\documentclass[12pt,preprint]{aastex}

\newcommand{\msun}{$M_{\odot}$}
\newcommand{\vmon}{A0620-00}
\newcommand{\flam}{ergs~cm$^{-2}$~s$^{-1}$~\AA$^{-1}$}

\begin{document}
\title{Multiwavelength Observations of \vmon\ in Quiescence}

\author{Cynthia S. Froning}
\email{cynthia.froning@colorado.edu}
\affil{Center for Astrophysics and Space Astronomy, University of Colorado, \\
  593 UCB, Boulder, CO 80309-0593\altaffilmark{1}}
  
\author{Andrew G. Cantrell}
\email{andrew.cantrell@yale.edu}
\affil{Department of Astronomy, Yale University, P.O. Box 208101, New Haven, CT  06520} 

\author{Thomas J. Maccarone}
\email{t.j.maccarone@soton.ac.uk}
\affil{School of Physics and Astronomy, University of Southampton, SO17 1BJ, UK}
        
\author{Kevin France, Juthika Khargharia, Lisa M. Winter\altaffilmark{2}}
\email{kevin.france@colorado.edu,juthika.khargharia@colorado.edu,lisa.winter@colorado.edu}
\affil{Center for Astrophysics and Space Astronomy, University of Colorado, \\
  593 UCB, Boulder, CO 80309-0593}
        
\author{Edward L. Robinson}
\email{elr@astro.as.utexas.edu}
\affil{Department of Astronomy, University of Texas at Austin, 
Austin, TX 78712}

\author{Robert I. Hynes}
\email{rih@theory.phys.lsu.edu}
\affil{Department of Physics and Astronomy, Louisiana State University, 
Baton Rouge, LA 70803}

\author{Jess W. Broderick}
\email{J.Broderick@soton.ac.uk}
\affil{School of Physics and Astronomy, University of Southampton, SO17 1BJ, UK}

\author{Sera Markoff}
\email{s.b.markoff@uva.nl}
\affil{Astronomical Institute `Anton Pannekoek', University of Amsterdam, Science Park 904, 1098 XH, the Netherlands}

\author{Manuel~A.~P.\ Torres, Michael Garcia}
\email{mtorres@head.cfa.harvard.edu, garcia@head.cfa.harvard.edu}
\affil{Harvard-Smithsonial Center for Astrophysics, 60 Garden St., Cambridge, MA 02138\altaffilmark{3}}

\author{Charles D. Bailyn}
\email{charles.bailyn@yale.edu}
\affil{Department of Astronomy, Yale University, P.O. Box 208101, New Haven, CT  06520} 

\author{J. Xavier Prochaska}
\email{xavier@ucolick.org}
\affil{Department of Astronomy and Astrophysics, UCO/Lick Observatory, University of California, 1156 High Street, Santa Cruz, CA 95064}

\author{Jessica Werk}
\email{jwerk@umich.edu}
\affil{University of Michigan, 500 Church St., Ann Arbor, MI 48109}

\author{Chris Thom}
\email{cthom@stsci.edu}
\affil{Space Telescope Science Institute, 4700 San Martin Dr., Baltimore, MD 21218}

\author{St\'{e}phane B\'{e}land, Charles W. Danforth, Brian Keeney}
\email{stephane.beland@colorado.edu, charles.danforth@colorado.edu, brian.keeney@colorado.edu}
\affil{Center for Astrophysics and Space Astronomy, University of Colorado, \\
  593 UCB, Boulder, CO 80309-0593}
    
\and
\author{James C. Green}
\email{james.green@colorado.edu}
\affil{Center for Astrophysics and Space Astronomy, University of Colorado, 593 UCB, Boulder, CO, 80309-0593\altaffilmark{1}}
\altaffiltext{1}{Department of Astrophysical and Planetary Sciences, University of Colorado}
\altaffiltext{2}{Hubble Fellow.}
\altaffiltext{3}{SRON, Netherlands Institute for Space Research, Sorbonnelaan 2, 3584 CA Utrecht, the Netherlands}

\begin{abstract}
We present contemporaneous X-ray, ultraviolet, optical, near-infrared, and radio observations of the black hole binary system, \vmon, acquired in 2010 March.  Using the Cosmic Origins Spectrograph on the Hubble Space Telescope, we have obtained the first FUV spectrum of \vmon, as well as NUV observations with STIS.  The observed spectrum is flat in the FUV and very faint (with continuum fluxes $\simeq1e-17$~\flam). The UV spectra also show strong, broad (FWHM$\sim$2000~km~s$^{-1}$) emission lines of \ion{Si}{4}, \ion{C}{4}, \ion{He}{2}, \ion{Fe}{2}, and \ion{Mg}{2}. The \ion{C}{4} doublet is anomalously weak compared to the other lines, which is consistent with the low carbon abundance seen in NIR spectra of the source. Comparison of these observations with previous NUV spectra of \vmon\ show that the UV flux has varied by factors of 2--8 over several years.  We compiled the dereddened, broadband spectral energy distribution of \vmon\ and compared it to previous SEDs as well as theoretical models. The SEDs show that the source varies at all wavelengths for which we have multiple samples. Contrary to previous observations, the optical-UV spectrum does not continue to drop to shorter wavelengths, but instead shows a recovery and an increasingly blue spectrum in the FUV.  We created an optical-UV spectrum of \vmon\ with the donor star contribution removed.  The non-stellar spectrum peaks at $\simeq$3000~\AA. The peak can be fit with a T=10,000~K blackbody with a small emitting area, probably originating in the hot spot where the accretion stream impacts the outer disk. However,  one or more components in addition to the blackbody are needed to fit the FUV upturn and the red optical fluxes in the optical-UV spectrum.  By comparing the mass accretion rate determined from the hot spot luminosity to the mean accretion rate inferred from the outburst history, we find that the latter is an order of magnitude smaller than the former, indicating that $\sim$90\% of the accreted mass must be lost from the system if the predictions of the disk instability model and the estimated interoutburst interval are correct.  The mass accretion rate at the hot spot is $10^{5}$ the accretion rate at the black hole inferred from the X-ray luminosity.  To reconcile these requires that outflows carry away virtually all of the accreted mass, a very low rate of mass transfer from the outer cold disk into the inner hot region, and/or radiatively inefficient accretion. We compared our broadband SED to two models of \vmon\ in quiescence, the ADAF model and the maximally-dominated jet model.  The comparison suggests that strong outflows may be present in the system, indicated by the discrepancies in accretion rates and the FUV upturn in flux in the SED. 

\end{abstract}


\section{Introduction} \label{sec_intro}

Low mass X-ray binaries (LMXBs) are interacting binary systems in which a late-type star fills its Roche lobe and transfers material via an accretion disk to a neutron star or black hole accretor.   LMXBs have proven to be effective test beds for the study of the physics of accretion and probes of fundamental physics in the strong gravity regime.  For example, observations of LMXBs have been used to determine the geometry of accretion flows \citep{hynes2010},  demonstrate evidence for black hole event horizons \citep{garcia2001}, find black hole and neutron star masses \citep{farr2010,ozel2010,charles2006}, measure black hole spins \citep{mcclintock2010,miller2009}, track thermonuclear bursts on neutron star surfaces \citep{strohmayer2006}, and probe disk-jet and black hole-jet coupling processes \citep{fender2010}.  Radio and X-ray emission in sub-Eddington black hole systems are correlated, with a correlation with black hole mass that extends over eight orders of magnitude from stellar mass LMXBs to AGN \citep{merloni2003,falcke2004}. This correlation, dubbed the ``Fundamental Plane'' of black hole activity, establishes the presence of common accretion processes at work across all observed mass scales.  These fundamental physical processes are often best studied in LMXBs, where the shorter variability time scale allows for extensive tracking and modeling of transient phenomena and the systems are not obscured or confused by emission from the AGN host galaxy. 

The observational properties of steady-state accretion disks are generally well described by multi-temperature thermal emission from a classic thin disk, where the disk temperature varies radially as T(R) $\propto$ R$^{-3/4}$ \citep{shakura1973}.  In systems in the ``low/hard state'' or in quiescence, where the accretion rate is roughly a few percent of the Eddington luminosity, the picture is less well defined \citep[see discussion in][]{mcclintock2006}.  The X-ray spectrum is dominated by hard, non-thermal emission well-described by a power-law but inconsistent with thermal emission from the disk. This emission is generally ascribed to thermal Comptonization in a hot, optically thin accretion disk corona located near the center of the system.  A physical model for this corona was developed by \citet{narayan1994,narayan1995} who proposed that the thin disk is disrupted at large radii, forming an advection-dominated accretion flow (ADAF) near the center of the system. The ADAF is a radiatively inefficient flow in which most of the energy released by viscous dissipation is directly advected with the flow.  \citet{merloni2001a,merloni2001b} propose an alternate model in which the accretion disk corona is driven by magnetic flares from the underlying accretion disk.  These and other publications on radiatively inefficient accretion flows and accretion disk coronae abound, but the fundamental structure of the disk/corona at low accretion rates remains an unresolved problem.

In recent years, moreover, there has been an increasing awareness of the ubiquity of outflows in LMXBs across the full range of observed accretion states.  The original ADAF papers discussed the likelihood of outflows associated with the ADAF in quiescent systems, which was later expanded upon by \citet{blandford1999} to propose a model in which most of the accreting gas is driven from the system by strong winds.  Even more dramatic has been the explosion of interest in jet emission in LMXBs \citep[see][and sources therein]{fender2006}. Flat or inverted spectra have been observed in the radio to millimeter regime in several LMXBs in the low/hard state. The spectra are attributed to synchrotron emission from a highly collimated outflow.  The jet emission may not be restricted to the radio, however: it has been proposed that the jet can dominate the broadband spectrum of quiescent LMXBs from the radio to  X-rays \citep[e.g.,][]{markoff2001}.  

One of the best-studied LMXBs is the black hole system, \vmon.  \vmon\ was discovered when it went into outburst in 1975 \citep{elvis1975}.  \vmon\ has been in a quiescent state since 1976, during which extensive observations have shown that the system is composed of a K-type donor star transferring mass to a  black hole via an accretion disk \citep{oke1977,mcclintock1983,mcclintock1986}.   The black hole mass in \vmon\ has been precisely determined, M$_{BH}$ = 6.6$\pm$0.25~\msun\ and the continuum fitting method for estimating black hole spins from the thermal emission from the accretion disk in the soft state gives an estimated spin of a$_{*}$ = 0.12$\pm$0.19 \citep{cantrell2010, gou2010}. Cantrell et al.\ also determined the distance to \vmon\ as $d=1.06\pm0.12$~kpc. 

\citet{gallo2006} presented radio observations of \vmon, the first radio detection of a quiescent black hole binary, and one that extended the Fundamental Plane to black hole systems with luminosities as low as 10$^{-8.5}$ of the Eddington luminosity.   \vmon\ was also detected by Spitzer at 24$\mu$m, which \citet{muno2006} attributed to thermal emission from a circumbinary disk.   \citet{gallo2007}, however, noting the flat spectral index between the radio and the mid-IR,  argued that synchrotron emission from a jet was responsible for the emission in both bands.  They fit the radio to X-ray spectral energy distribution (SED) with a maximally jet-dominated model in which, aside from visible and near-infrared contributions from the donor star and the disk, the SED is dominated at all energies by emission from the jet.  This result is in contrast to ADAF models of the quiescent disk in \vmon\ in which the X-ray emission is dominated by the ADAF and the visible-UV emission by the outer thin disk \citep{narayan1996,narayan1997}. Hence, \vmon\ is one of the key systems for which extensive multiwavelength observations have allowed for tests of multiple quiescent accretion models, from which has sprung general consensus about the importance of non-thermal emission from the jet and disk corona but lingering disagreements about specific questions concerning the energetics of the corona, the structure of the inner disk, and the relative contributions of each component in different wavebands \citep{narayan2008,markoff2010}.
 
Here, we present UV spectroscopy of \vmon\ obtained with the Cosmic Origins Spectrograph (COS) and Space Telescope Imaging Spectrograph (STIS) on the Hubble Space Telescope (HST). The COS spectra are the first far-ultraviolet (FUV; $\lambda <2000$~\AA) observations of this faint source. The UV wavebands provide several key tracers of the structure of LMXB accretion disks and outflows, including line emission from the accretion disk chromospheres and disk winds \citep[e.g.,][]{bayless2010,haswell2002} and probes of the SED near the expected peak temperatures of thermal emission components in the disk. As a result, the FUV can provide key constraints on disk models in LMXBs \citep{hynes2009}. We combine the UV data with X-ray, optical, near-infrared, and radio observations, all acquired over a four day period, to create a broadband SED.   Using the multiwavelength data set, we examine changes in the UV flux and broadband SED in the system over time and compare the current properties of the system to models of the structure of quiescent black hole X-ray binaries. 

\section{Observations and Data Reduction} \label{sec_observ}

\subsection{HST UV Spectroscopy}

\vmon\ was observed with COS on HST on 23 March 2010. The total exposure time was 4.6~hr spread over a clock time of about 9 hours. A summary of all the observations can be found in Table~\ref{tab_obs}. We observed \vmon\ using the FUV G140L grating in the 1230~\AA\ setting, which covers $\sim$1300 -- 2400~\AA\ on the A segment of the FUV detector at a spectral resolution of $\Delta \lambda \sim 0.5$~\AA\ (however, instrument sensitivity is very low for $\lambda > 1800$\AA).  In the same setting the B segment covers $\sim$200 -- 1170~\AA\ with appreciable sensitivity down to the Lyman limit  \citep{mccandliss2010}.  We stepped the grating to different fp-pos positions for each exposure to minimize the effects of fixed pattern noise. Additional information about the design and on-orbit performance of COS can be found in  \citet{osterman2011} and the COS Instrument Handbook \citep{dixon2010}. 

We retrieved the COS data from the Multi-Mission Archive at STScI (MAST).  The data had been processed with V. 2.11b of CALCOS. That version of CALCOS did not correctly perform the wavelength and flux calibration for Segment B data in G140L, so we re-processed the Segment B data using a newer version of the CALCOS pipeline (v2.12), employing specially-created reference files for flux and wavelength calibrations in the short-wavelength segment B setting. The custom reference files were created to supplement the CALCOS 2.12 release that included a first order dispersion solution for $\lambda < 1150$~\AA\ and a flux calibration curve.  A detailed discussion of the development of the calibration files is available in \citet{shull2010}, based on data first presented in \citet{mccandliss2010}.
The absolute flux accuracy for the Segment B data presented here is about 10\%--15\%. We coadded the output one-dimensional spectral data products using a custom IDL code, described in \citet{danforth2010}\footnote{See also \url{http://casa.colorado.edu/~danforth/science/cos/costools.html}}.  The code performs a partial flat field correction (removing shadows cast by the detector ion repeller grid), combines different wavelength settings, and creates a weighted mean spectrum (with lower weight given to regions of uncertain flux calibration near detector edges). 

We also observed \vmon\ with STIS on 24 March 2010 using the G230L grating and the 52$\times$0.5 slit, which covers 1507--3180~\AA\ at $\Delta \lambda$=3.2~\AA.  The total exposure time was 2.9 hr acquired over a $\sim$7 hr time period. We retrieved the data from MAST.  We also retrieved the 1998 STIS observations of \vmon\ (program GO-7393). The 2010 data were processed by CALSTIS V. 2.26 and the 1998 data by CALSTIS V.2.23. Since STIS was repaired during Servicing Mission Four, the NUV MAMA detector has been showing elevated dark count rates (see the STIS Instrument Handbook for more information; Proffitt et al. 2010). Although the dark count rate has been declining, in early 2010 it was still at $\sim$0.005 counts~sec$^{-1}$~pixel$^{-1}$, a factor of 4 above pre-failure levels (STScI Analysis Newsletter, February 2010). Probably as a result of the noisier data, CALSTIS was unable to automatically extract the one-dimensional spectrum from the two-dimensional spectral image in the 2010 observations, so we extracted the spectra using the IRAF\footnote{"IRAF is distributed by the National Optical Astronomy Observatory, which is operated by the Association of Universities for Research in Astronomy (AURA) under cooperative agreement with the National Science Foundation.}/STSDAS task x1d, providing the location of the cross-dispersion profile in each exposure from inspection of the two-dimensional images. 

Figures~\ref{hst_cos} and~\ref{hst_stis} show the time-averaged UV spectra from the 2010 COS and STIS observations of \vmon, respectively.  The red error bars in the figures show the Poisson noise from CALCOS and CALSTIS, propagated through the averaging performed by the coaddition tool and the pixel binning.  While these error bars are roughly representative of the noise in our spectra, they do not give the true uncertainty, because the data reduction pipelines do not properly handle Poisson noise in the low count rate regime.  In particular, the pipelines incorrectly assign zero variance to zero count events and do not adopt two-sided confidence limits to take into account the zero probability of negative net counts in background-subtracted spectra. Accordingly, we present these error bars to give a visual estimate of the scatter in each bin and between bins, but do not use them in our analysis (except as input to specfit when fitting the emission lines).   

For our spectral energy distribution, we calculate the mean in several line-free spectral regions (shown in purple in the figures) and use the uncertainty on the mean for our error bars. The error bars do not incorporate uncertainties in absolute flux calibration or in background subtraction, however, that could move all the UV data relative to our other wavebands. The absolute flux calibration uncertainty is fairly low for both instruments, $\sim$5\% \citep{dixon2010,proffitt2010}.  Our target was well-centered in the COS aperture, so we did not experience any vignetting losses.  The COS FUV channel has shown evidence of on-orbit sensitivity degradations which are still being characterized and may account for an additional uncertainty in the absolute flux calibration of $\leq$4\% \citep{osten2010}.

\subsection{Swift X-ray and UV Imaging}

Swift made pointed observations of \vmon\ several times between 22 and 25 March 2010.  The observations for the different dates have Swift identifiers 00031635001, 00031635002, 00031635003, and 00031635004, respectively. We retrieved and calibrated the X-Ray Telescope (XRT; Burrows et al. 2005) and UV/Optical Telescope (UVOT; Roming et al. 2005) data.  Observation data and cumulative exposure times are listed in Table~\ref{tab_obs}.  

Because the individual XRT observations all have poor signal to noise, we add the four dates' data together for a total exposure time of 19939 sec.  We use a 20 arcsecond source circle.  This corresponds to an encircled energy fraction of 75\%, and is chosen as a compromise between including all the source photons and minimizing background.  \vmon\ is in a relatively uncrowded X-ray field. There is one source located 10$\arcsec$ away, but it is 5\%--10\% as bright as \vmon\ (based on examination of archival Chandra observations of the field) and therefore likely contributes $<$10\% of the counts in the \vmon\ extraction circle.  We find a total of 25 photons within the source regions.  We estimate the background from off-source regions with 100-114 (236$\arcsec$--269$\arcsec$) pixel radii, depending on which image is used.  We find that there is a total of $2.9\pm0.1$ photons per 20$\arcsec$ radius circle.  We thus estimate $22.1\pm5.0$ source counts within the 20$\arcsec$ radius, with the uncertainties dominated by Poisson statistics of the source counts.  Extrapolating to the full point spread function, and dividing by the exposure time gives $1.5\pm{0.3}\times10^{-3}$ counts~sec$^{-1}$.

There are too few counts for X-ray spectroscopy, so we converted the count rate into a flux using W3PIMMS\footnote{\url{http://heasarc.gsfc.nasa.gov/Tools/w3pimms.html}}.  We used $N_H$=1.6$\times10^{21}$ \citep{mcclintock1995,gallo2006}, and tried several different power law spectral models.  For $\Gamma=2.0$, the 0.5--8.0 keV unabsorbed flux is $6.4\times10^{-14}$ ergs~sec$^{-1}$~cm$^{-2}$, while the same quantity is $7.0\times10^{-14}$
ergs~sec$^{-1}$~cm$^{-2}$ with $\Gamma=1.7$ and $5.8\times10^{-14}$
ergs~sec$^{-1}$~cm$^{-2}$ with $\Gamma=2.5$.  Previous Chandra observations of \vmon\ have found values of $\Gamma$ from 2.06 \citep{gallo2006} to 2.26 \citep{mcclintock2003}. For our SED (Table~\ref{tab_sed}) we use the flux for the $\Gamma$=2.0 fit to allow a direct comparison to the \citet{gallo2006} SED. The statistical errors are, as stated above, about 20\%, while the systematic errors due to uncertainty in the spectral shape are likely to be about 10\%.  The source flux is thus consistent with that found in August of 2005 by \citet{gallo2006}, and a factor of about 2 larger than that found in February of 2000 by \citet{kong2002}.

For the Swift UVOT data, we used the archived level 2 processed data files.  A source region was defined with a 10$\arcsec$ circular radius centered on the position of \vmon.  A background region was defined with a 20$\arcsec$ circular aperture in a region free from additional sources, near \vmon.  Using these defined regions and the 
Swift Ftool\footnote{See  \url{http://heasarc.gsfc.nasa.gov/docs/software/ftools/ftools\_menu.html} } 
uvotsource, we extracted the background-subtracted flux from the Swift UVOT observations in the UVW1, UVW2, UVM1, and U filters.  

\subsection{SMARTS/ANDICAM Optical and Near-Infrared Imaging}

We observed \vmon\ at visible and near-infrared wavelengths using the ANDICAM instrument mounted on the 1.3-m telescope at CTIO.  ANDICAM is operated by the Small and Moderate Aperture Research Telescope Systems (SMARTS) consortium\footnote{See \url{http://www.astro.yale.edu/smarts}}.  We obtained BVRIJHK observations of \vmon\ nightly (weather permitting) from 18 -- 31 March 2010, spanning several days around the HST observations.  The BVI exposures are 6 minutes each.  The JHK exposures each consist of eight dithered 30-second exposures, which are sky-subtracted and then combined.  The data were reduced using standard IRAF tasks for calibration and photometry. We determined the flux calibration and photometric errors using comparison stars. The BVI data are calibrated using Landolt standards in other fields, while the JHK magnitudes are calibrated using the 2MASS magnitudes of field stars. \vmon\ has been monitored regularly by the SMARTS consortium for over a decade.  Further details on how the data acquisition, reduction, and photometric calibration are undertaken for these observations are available in \citet{cantrell2008}.


\subsection{ATCA Radio Data}

We observed A0620$-$00 with the Australia Telescope Compact Array (ATCA) on both 2010 March 23 and 24. Simultaneous 5.5 and 9.0 GHz observations were conducted with the Compact Array Broadband Backend (CABB); the bandwidth is about 2 GHz at each frequency. Because of the declination of the target, we used the hybrid H168 array, which includes a north-south spur to enable reasonable coverage of the $uv$-plane. The total integration time on-source was 7.0 hours, split approximately evenly between the two observing sessions. The primary calibrator was B1934$-$638, and the secondary calibrator was B0639$-$032.  

We reduced and imaged the data with {\sc MIRIAD} \citep*[][]{sault95}; note that a single image was made at each frequency using the full 7.0 hour dataset. After flagging, the effective frequencies of the two bands are 5.48 and 9.00 GHz. We set Briggs' robust weighting parameter \citep[][]{briggs95} to 0.5 when forming the 9.00 GHz map, but found that a value of 0.0 at 5.48 GHz gave the best compromise between sensitivity and the suppression of sidelobes from nearby sources. As A0620$-$00 is very close to the celestial equator, it was necessary to use SIN (sine) projection when forming the images; though initially we had to modify the {\sc MIRIAD} source code to do this, the {\sc MIRIAD} task {\sc INVERT} has since been updated with a built-in option. In addition, because of the wide bandwidths, we used the multi-frequency deconvolution algorithm {\sc MFCLEAN} \citep[][]{sault94}. The angular resolutions are 38 arcsec $\times$ 1.6 arcsec (beam position angle -1.2$^{\circ}$) and 23 arcsec $\times$ 1.3 arcsec (position angle -1.1$^{\circ}$) at 5.48 and 9.00 GHz, respectively. At 5.48 GHz, the rms noise level is 13.5 $\mu$Jy beam$^{-1}$, while at 9.00 GHz it is 16 $\mu$Jy beam$^{-1}$. 

The source is not detected at either frequency. The 5$\sigma$ upper limits are therefore 67.5 $\mu$Jy beam$^{-1}$ and 80 $\mu$Jy beam$^{-1}$ at 5.48 and 9.00 GHz, respectively.  

\subsection{Keck Optical Spectroscopy}

Using the Low Resolution Imaging Spectrometer \citep[LRIS;][]{oke1995} at the Keck Observatory, we obtained visible band spectra of \vmon\ on 2010 Mar 25. The instrument was configured using dichroic D560 coupled with the 600/7500 grating on the red arm and the 600/4000 grism on the blue arm. The data were taken through the 1$\arcsec$ long slit.  In this configuration, LRIS provides coverage from 3010--5600~\AA\ at 3.8--4.1~\AA\ ($\sim$280~km~s$^{-1}$) resolution in the blue arm and 5600--8870 at 4.7~\AA\ ($\sim$160~km~s$^{-1}$) from the red arm. Data were calibrated using the LowRedux software package.\footnote{\url{http://www.ucolick.org/~xavier/LowRedux/index.html}} We calibrated the exposures with spectra of arc line emission lamps and fluxed the data with a sensitivity function derived from observations of G191B2B taken that night. The flux calibration does not include a precise estimate of slit-loss and therefore only provides an accurate estimate of the relative flux.

\section{Analysis} \label{sec_analysis}

\subsection{Ultraviolet Spectra}

\begin{figure}
\includegraphics[angle=90,scale=0.7]{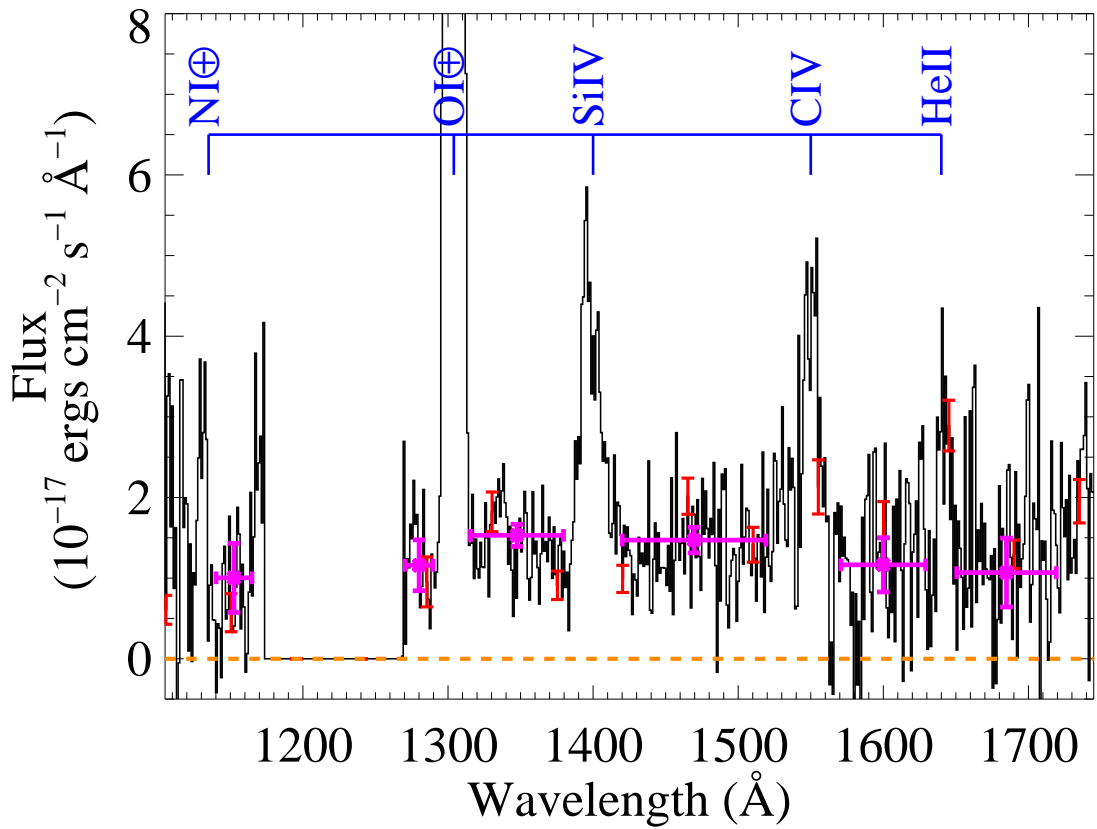}
\figcaption[fig1.ps]{The time-averaged COS FUV spectrum of \vmon\ is shown in black.  The observations were acquired on 23 March 2010. The spectrum has been binned to 2 resolution elements (15 pixel binning). The error bars shown in red are the statistical uncertainties from CALCOS propagated through the binning of the data points. Prominent emission features are labeled in blue (with airglow lines labeled with the circled plus signs).  Finally, the purple bars show mean continuum fluxes, with the horizontal bars indicating the range over which the mean was calculated and the vertical bars showing the uncertainty of the mean.\label{hst_cos}}
\end{figure}

\begin{figure}
\includegraphics[angle=90,scale=0.7]{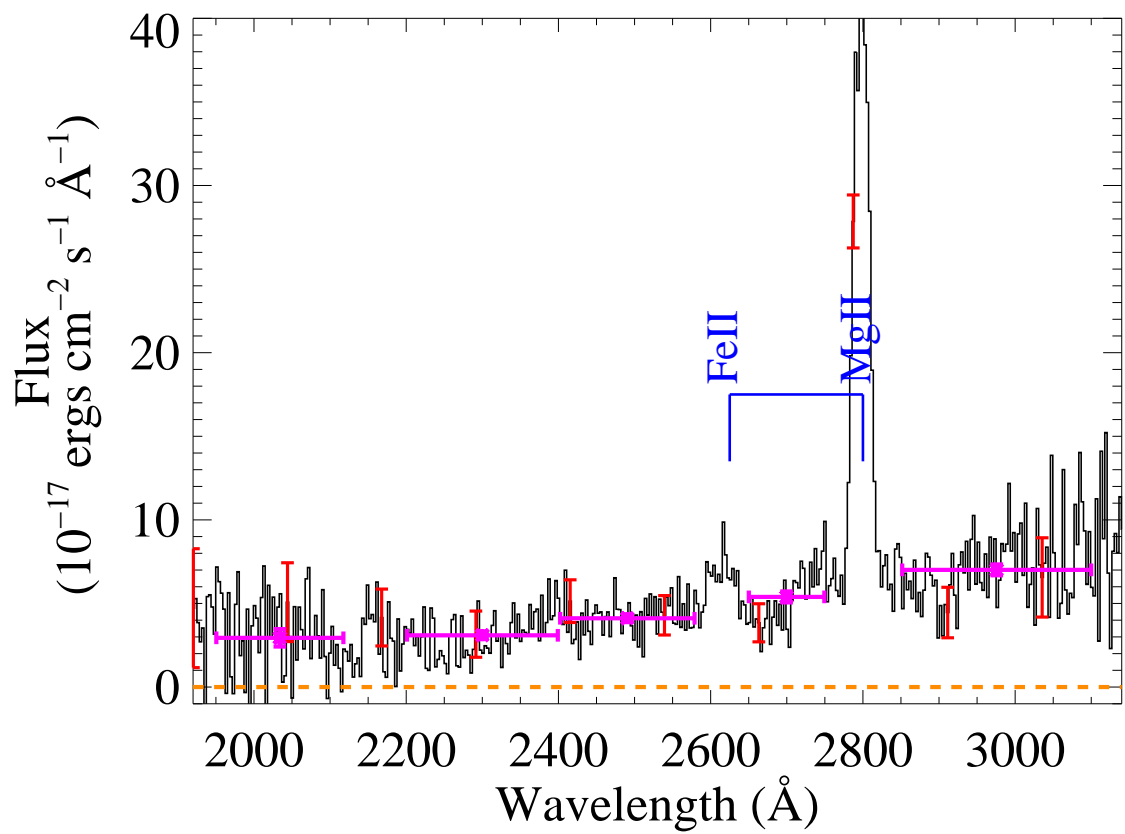}
\figcaption[fig2.ps]{The time-averaged STIS NUV spectrum of \vmon, acquired on 24 March 2010. The spectrum has been binned by two pixels (one resolution element). \label{hst_stis}}
\end{figure}

The UV spectra of \vmon\ are shown in Figures~\ref{hst_cos} and~\ref{hst_stis}. The spectra have not been corrected for interstellar reddening.  The observed continuum is flat in the FUV (1150--1700~\AA) and red in the NUV (1800--3200~\AA), with fluxes in the FUV $\simeq1\times10^{-17}$~\flam\ and 3--7$\times10^{-17}$~\flam\ in the NUV. The COS FUV spectrum shows prominent, broad (FWHM$\sim$2000~km~s$^{-1}$) emission lines of \ion{Si}{4}, \ion{C}{4}, and \ion{He}{2}. In the NUV, the STIS spectra show emission lines of \ion{Fe}{2} and \ion{Mg}{2}. Integrated line fluxes are given in Table~\ref{tab_lines}, based on Gaussian fits to the observed spectra using Specfit \citep{kriss1994}.  The \ion{Si}{4} line was slightly better fit with the doublet fixed to a 2:1 ratio than 1:1 but the difference was not statistically significant.  For \ion{C}{4}, we could not distinguish between fits with 2:1 or 1:1 line ratios; we give the latter in Table~\ref{tab_lines}. The line centroids were within $\sim$200~km~s$^{-1}$ of their rest velocities.

Average continuum fluxes for the line-free regions (shown in purple in Figures~\ref{hst_cos} and~\ref{hst_stis}) are given in Table~\ref{tab_cont}.  The table also contains the UV measurements made by Swift/UVOT.  Overall, the UVOT fluxes are brighter than the STIS fluxes.  Most of the UVOT UV filters (particularly UVW2 and UVW1) have substantial red leak which accounts for much of the discrepancy between the measurements.  The UVOT U filter does have a square response profile, although it has a broader high throughput region than that of Johnson U. The UVM2 filter has the most ``UV-pure" filter coverage, so we compared the STIS fluxes (acquired on March 24) to the UVM2 data (acquired on March 23, the same day as the COS observations) by creating an average flux from the STIS spectrum weighted by the UVM2 filter profile.  The weighted STIS flux is $3.4\pm 0.2 \times 10^{-17}$~\flam\ while the UVM2 flux is $5.0\pm 1.7 \times 10^{-17}$~\flam.  The UVM2 flux was calculated using the count rate to flux density conversion factor given in Table 9 of \citet{poole2008}, which uses stellar spectra (rather than the default gamma-ray burst model spectra) to determine the conversion factor.  The error bars do not include the absolute flux calibration uncertainties which are, however, small:  5\% for STIS and 2.8\% for the UVOT UV filters \citep{proffitt2010,poole2008}.   

A comparison of the STIS and UVM2 fluxes shows that \vmon\ may have varied in the NUV by 50\% in the one day between observations, being brighter during the COS observations.  However, given the uncertainties on the UVOT measurement, the data are also consistent with no variation. We consider the 50\% variation as a rough upper limit on changes in the UV flux in \vmon\ between our observations.  

\begin{figure}
\plotone{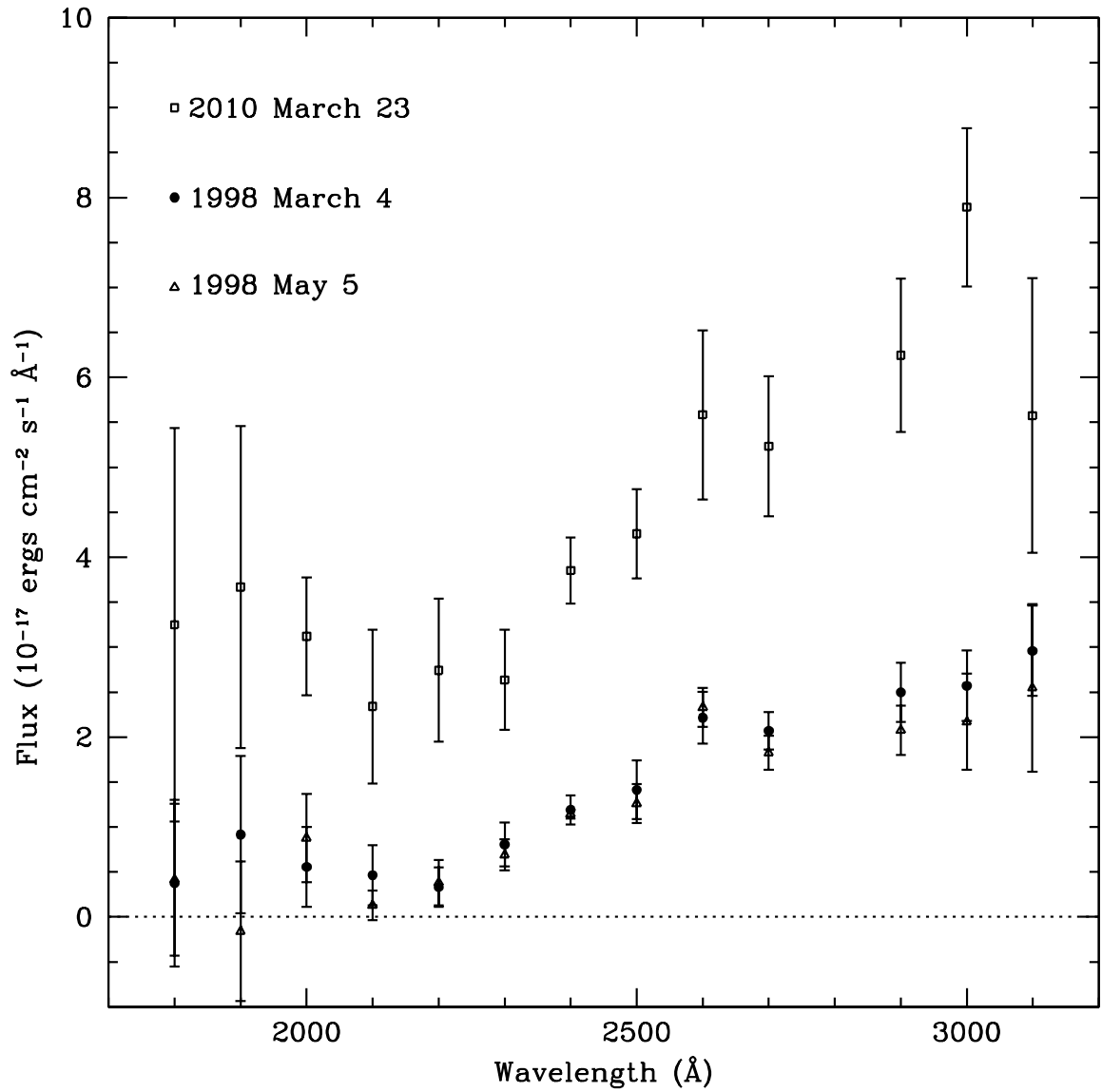}
\figcaption[fig3.ps]{Comparison of the three STIS NUV observations  of A0620--00. The fluxes have not been corrected for reddening. Each point shows the mean continuum flux over a 100~\AA\ bin. The error bars are the rms scatter about the mean. The \ion{Mg}{2} $\lambda$2800 line is not shown.\label{varib_stis}}
\end{figure}

\vmon\ has now been observed three times by STIS.  Following the treatment of \citet[][hereafter MR]{mcclintock2000} in their Figure 2, we compare the STIS spectra of \vmon\ to search for long-term variabilty. Figure~\ref{varib_stis} shows the continuum fluxes obtained during each of the three STIS NUV observations of \vmon. Each point is the average of a 100~\AA\ spectral region, with the error bars showing the standard deviation of the input points about the mean in each bin. Note that the fluxes from the 1998 spectra that we present are similar to but not the same as those presented by \citet{mcclintock2000}.  Changes in the CALSTIS pipeline were made between 2000 and 2004 to improve both the flux calibration (including implementation of time-dependent calibration files) and background subtraction algorithms, resulting in slightly different calibrated spectra (C. Proffitt, private communication).  Our measurements and those of MR agree within the errors but in the recalibrated spectra the evidence of variability between the two 1998 observations seen by MR is less evident, particularly given the scatter in each bin. However, the 1998 March measurements remain brighter than the May ones, particularly for wavelengths $>$2800~\AA, where the former is about 20\% brighter than the latter, consistent with the MR estimates of the amplitude of the variability, $\sim$25\%. 

The 2010 NUV spectrum is substantially brighter than the 1998 spectra.  Because of the enhanced detector background emission, the uncertainties are larger in 2010 but even within those uncertainties, \vmon\ is brighter than in 1998 across the spectrum. The typical increases in flux are by
by factors of $\simeq$2--8. For reasons of clarity, the bin containing \ion{Mg}{2} is not shown in Figure~\ref{varib_stis}, but the \ion{ Mg}{2} integrated line flux is also a factor of 2.2--2.3 higher in 2010 than in the 1998 observations, consistent with the increase in the continuum level and representing a slight decrease in equivalent width (from to 179$\pm$6~\AA\ to 154$\pm$4~\AA).  

In 2003, \vmon\ transitioned from spending most of its time in a ``passive'' state to one in which it is typically ``active'' \citep{cantrell2008}. In its active state, \vmon\ is brighter and more variable than in its passive state: measured I-band magnitudes of \vmon\ span 0.4 mag in the active state versus 0.06 mag in the passive state.  Assuming that the donor star accounts for $\sim$75\% of the I-band flux \citep{cantrell2010} and that all of the 0.4 mag variability is in the remaining 25\% flux component, the second component is changing in I by a factor of 2.6.  Consequently, it is unsurprising to see variations of a factor of $\geq$2 in the UV (where the donor star contribution is negligible) between spectra taken in the passive state and in the active state.  That having been said, it is possible that the UV variability in \vmon\ is large even when the target is mostly passive.  MR noted that the STIS spectrum was half as bright as the FOS observations of \vmon\ acquired six years earlier \citep{mcclintock1995}. Thus, the FOS fluxes are closer to the 2010 STIS values, although the STIS observations are still a factor of 1.5--2 brighter. However, MR also caution that the FOS data were obtained pre-COSTAR and may suffer from uncertain flux calibration and background subtraction (the FOS prism and grating fluxes differ by $\sim$35\% in their region of overlap, for example).

Finally, using the time-tag event list, we searched for source variability in the COS exposures, binning the A segment data (away from geocoronal lines) into 120 sec bins.  The FUV count rate for \vmon\ is quite low: $\sim$0.5 counts~sec$^{-1}$ from the target over the 1260--1700~\AA\ bandpass, compared to $\sim$1.5~counts~sec$^{-1}$ background counts over the same waveband (based on the average of two detector regions above and below the target spectral extraction window).  The spectral count rate was very steady over the 9 hour observation period, with no exposure average varying from the total average count rate of 0.44$\pm$0.11 cps by more than 15\%, or less than the uncertainties from Poisson noise.

\subsection{Interstellar Reddening}

\citet{hynes2005a} reviewed various determinations of the interstellar reddening along the line of sight to \vmon\ and concluded that the most robust measurement is that of \citet{wu1983}, who obtained E(B--V) = 0.35$\pm$0.02.  The Wu et al.\ measurement was based on fits to the 2175~\AA\ interstellar absorption feature obtained by the five-channel spectrophotometer aboard the Astronomical Netherlands Satellite when \vmon\ was in outburst in 1975.  \citet{cantrell2010} found comparable reddening values for \vmon:  E(B--V) = 0.30$\pm$0.02 based on the stellar colors and assuming a K5V donor star. The near agreement suggests that the reddening law is fairly standard from IR to UV wavelengths. MR fit the 2175~\AA\  feature in their quiescent STIS spectra of \vmon\ but, due to the faintness of the source, were unable to constrain the reddening to better than $0.3 \leq$ E(B--V) $\leq 0.7$. We examined whether we could improve upon this result using the 2010 STIS observations spectra.  Unfortunately, although \vmon\ was brighter in 2010, the enhanced STIS background resulted in a noisy spectrum that was no better, even when combined with the 1998 data, for constraining the reddening. Here, we adopt the Wu et al.\ value, E(B-V)=0.35 and R$_{V}$ = 3.1.

\subsection{The Spectral Energy Distribution} \label{sec_sed}

\begin{figure}
\includegraphics[angle=-90,scale=0.7]{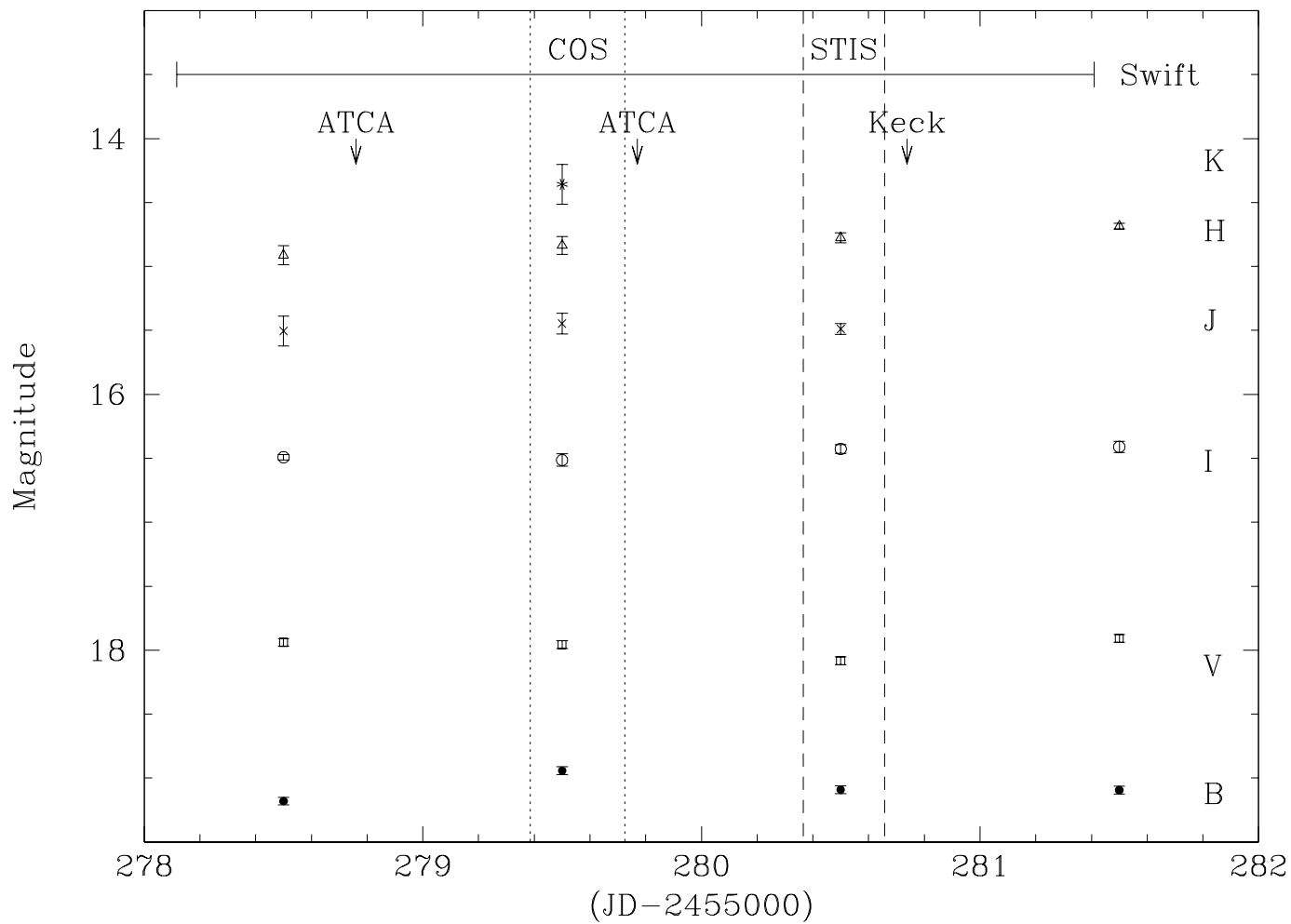}
\figcaption[fig4.ps]{The times of the multiwavelength observations of \vmon. The data points are BVIJHK observations from SMARTS. The error bars on the points are uncertainty on the differential photometry or the scatter between multiple observations on the same night, whichever is larger. The time of the COS observation is marked with dotted lines and the STIS observation with dashed lines. The  ATCA radio and Keck optical spectroscopy observations are marked with arrows. The Swift observation interval is labeled with the solid bar. \label{fig_time}}
\end{figure}

Figure~\ref{fig_time} shows the times of the \vmon\ observations relative to each other.  (We also have four more SMARTS BVIH observations, two before and two after the time interval shown in the figure.)  The Swift X-ray data were acquired over a four-day period, from which we obtain a single measurement.  The Swift U-band observations (the only UVOT data we use in the SED) were obtained starting a few hours after the STIS observations.  The COS and STIS UV spectra were acquired a day apart with the two radio observations (which were treated as a single dataset in the imaging process) spanning the COS data acquisition interval.  We obtained optical/NIR photometry nightly during this interval, including measurements made while the UV data were acquired.  These X-ray, UV, optical, NIR, and radio data allow us to construct a broadband SED of \vmon\ based on quasi-simultaneous (overlapping over a three day period) observations.  

The NIR magnitudes were fairly stable during this period. Over the two week interval, the I magnitudes varied by 0.12~mag while the H magnitudes changed by 0.2~mag.  Between the two days of the UV observations, the I-band flux brightened by 8\% and the H by 6\%, although in both cases the scatter within the individual observations in each night (partially due to periodic orbital variations) is comparable to the difference between the mean values from night to night. The two J observations were statistically indistinguishable. \vmon\ was more variable in B and V.  The B magnitudes varied by 0.44~mag over two weeks, though the variation was smaller (0.15~mag, or a 15\% decrease in flux) over the two nights when the COS and STIS observations were acquired.  In V, the magnitudes varied by up to 0.25~mag over two weeks and 0.13~mag (12\% decrease in flux) between the two UV observation intervals. However, the orbital phases covered by the single B and V observations on each night changed from $\Phi$ = 0.88--0.89 on 23 March to $\Phi$ = 0.98--0.99 on 24 March.  The light curve varies by $\sim$10\% over those phases simply due to the donor star modulation (e.g., the V light curve in Figure 2 of Cantrell et al. 2010).  As a result, the non-donor star variability in B and V may be as low as 2--5\%, and it is impossible to determine if this represents short-term flickering or a slower drift from one day to the next. 

For the spectral energy distribution (SED), we used the average of the four nights around our UV observations to determine the mean magnitudes for the optical/NIR.  We set the error bars to equal the rms scatter about the mean.  The resultant error bars are then combined in quadrature with the error bar on the absolute calibration for each filter to arrive at the final uncertainties. (Note that for J, we have three nights of data and for K just one night.)  We chose to average over a four-night period rather than just using the data on the nights of our UV observations because the scatter within the observations each night suggests that our mean values may be biased by short-term variability that is not consistently sampled in each filter and each night.  For the UV, the error bars do not  include an estimate of the uncertainty induced by possible variability in the day between the COS and STIS data acquisition.  As noted above, a comparison between STIS and the Swift UVOT measurement taken on the same day as the COS observations place a rough upper limit of 50\% on the UV variability between the two days.

Figure~\ref{fig_sed} shows the broadband SED for \vmon. The data have been dereddened assuming E(B--V) = 0.35 \citep{wu1983} and using the extinction relation of \citet{cardelli1989}.  For the use of future modelers, we also include the observed (not corrected for interstellar extinction) fluxes in Table~\ref{tab_sed}.  On the same figure we include two previous SEDs for \vmon, taken from MR and \citet{gallo2007}.  The MR data include the 1998 March STIS NUV spectra combined with 1992 January optical/NUV FOS spectra.  We only show their points for $\lambda <$3500~\AA\ because at longer wavelengths they subtracted out the donor star contribution.  They used the same dereddening correction that we adopt here. The \citet{gallo2007} SED includes simultaneous radio and X-ray data, with the optical/NIR (IVH) data being acquired one day before (all in 2005 August).  The Spitzer IR data were acquired five months earlier, in 2005 March. Gallo et al.\ dereddened their data assuming E(B--V) = 0.39.  Here, we have adjusted their data to apply an E(B--V) = 0.35 dereddening to place all the observations on the same scale.

The broadband SED for \vmon\ clearly varies over time.  Although our 0.5--8.0~keV flux is the same as that of \citet{gallo2006,gallo2007}, their optical/NIR data is brighter, with fluxes higher by 20\% in V, 44\% in I, and 28\% in H.  As discussed earlier, our UV observations are substantially brighter than the previous observations presented in MR. Our STIS NUV fluxes are factors of 2--8 brighter than the 1998 data and a factor of 1.5--2 brighter than the 1992 FOS observations.  MR do not give the optical fluxes before subtracting the donor star, but Figure~4a of \citet{mcclintock1995} does show part of the UV/optical spectrum before subtraction for the FOS data.  From that figure, we can discern that the 2010 data is 25\% brighter in B than in 1992.

\subsection{The Donor Star Contribution and the Non-Stellar Spectrum}

The Keck spectrum was acquired about two hours after the STIS observation ended (and about five hours after the SMARTS observations on that night).  We used the Keck spectrum to determine the contribution of the donor star to the optical spectrum.  We compared the spectrum of \vmon\ to synthetic spectra compiled by \citet{munari2005} from Kurucz model atmospheres. The uniform dispersion (1~\AA/pixel) spectra were convolved with a Gaussian to match the resolution (R $\sim$ 1500) of the A0620-00 spectrum from LRIS-R. The template spectra were then broadened again to take into account the rotational velocity of the donor star star in A0620-00 \citep{marsh1994}.  The template spectrum was scaled and subtracted from the observed spectrum to determine the donor star fraction that minimized the residual.  (See \citet{froning2007} for more details on the method used.) 

In Figure~\ref{fig_dilute} we show the spectrum of \vmon\ around H$\alpha$ compared to a scaled, broadened synthetic spectrum (T=4500~K, $\log (g)$ = 4.5, solar abundance).  The donor star fraction found from this model is 58\%$\pm$6\%, where the uncertainty is given by the scatter in the best-fit fractions for different spectral lines in the $\sim$5600--6500~\AA\ spectral region.   We restricted our fits to this region as the longer wavelengths become increasingly contaminated by telluric absorption features.  We also fit the model spectra in the blue using the LRIS-B spectrum near H$\beta$ and confirmed that the fitted donor fraction at H$\beta$ (52\%$\pm$5\%) is consistent with the predicted donor fraction if the scaled spectrum from the red fit is extended into the blue. The donor star fraction (52\% near H$\beta$; 61\% for the 4750~K template) is comparable to the 44\% contribution at the same wavelengths found by \citet{neilsen2008} from 2006 observations. This has decreased compared to the 85\% donor star contribution found by \citet{marsh1994}, which is consistent with the increase in non-stellar emission in the system as \vmon\ transitioned to the more active state post-2003.

The donor star fraction depends on the adopted temperature of the template spectrum.  For a hotter template (T=5000~K), we obtain a donor star fraction of 76\%$\pm$3\% near H$\alpha$, while a 4750~K template gives a fraction of 67\%$\pm$6\%.   In general, the donor star spectral type in \vmon\ has been assumed to range from K4V to K7V, but \citet{gonzalez2004} found an earlier spectral type (T=4900~K, or roughly K3V) with supersolar abundances.  Here, we adopt the 4500~K template in accordance with previous NIR spectral fitting that rejected stars earlier than K5V as inconsistent with the broadband NIR SED and the H and K absorption spectra \citep{froning2007}. Although we prefer a later spectral type for reasons outlined in the 2007 paper, we caution that the true donor star temperature in \vmon\ is not definitively determined and a small mismatch between the temperature or metallicity of the template star and the true stellar values can lead to large errors in the derived donor fraction \citep{hynes2005}.   

In Figure~\ref{fig_disk}, we show the UV and optical dereddened spectra with the donor star contribution subtracted.  We flux-calibrated the Keck spectra using the I-band SMARTS photometric data acquired a few hours earlier.   The non-stellar spectrum has a blue peak near 3000~\AA\ and a possible secondary peak near 5000~\AA. We compared the non-stellar spectrum to a few simple prescriptions.  The upper panel of Figure~\ref{fig_disk} compares a 10,000~K blackbody spectrum to the non-stellar spectrum.  Note that because of the negligible contribution of the donor star to the UV spectrum, the flux peak near 3000~\AA\ is a robust result, independent of uncertainties in the true temperature of the donor star.  The blackbody curve was fit to the spectrum by eye; it is not a formal fit and is intended to be illustrative. Because of noise and gaps between the observed spectra, the blackbody peak is not tightly constrained. A temperature range of $\simeq$9000--11000~K provides equally good results:  $\gtrsim$11,000~K, the blackbody flux exceeds the FUV observed flux while $\lesssim$9000~K, the blackbody peaks moves too far to the red.  

A single blackbody does not describe the non-stellar spectrum. This remains true even if a different donor star temperature is used for the template spectrum:  the spectral shape is not that of a single blackbody (although for a hotter donor star and an increased stellar fraction, the non-stellar spectrum falls under the blackbody curve rather than exceeding it).  Given the apparent secondary peak in the spectrum near 5000~\AA, we also examined adding a second, cooler (T$\sim$5500~K) blackbody component.  While the addition of a second blackbody can give an improved fit to the long wavelength spectrum ($>$5000~\AA),  the same component also adds too much flux at shorter wavelengths causing the two-blackbody model to exceed the observed blue (3000--5000~\AA) flux. 

We also examined a blackbody plus single power law spectrum. The power law is of the form $f_{\lambda} \propto \lambda^{\alpha-2}$.  The two components were scaled to the observed spectrum at 3000~\AA\ and fit by eye.  The lower panel of Figure~\ref{fig_disk} compares a blackbody plus power law to the observed non-stellar spectrum.  The blackbody has a temperature, T$_{BB}$ = 10,000~K and the power law has a spectral index $\alpha$ = 1.9.  At 3000~\AA, the blackbody is scaled to 70\% of the observed flux and the power law to 30\%.  This simple model provides an adequate fit to the broadband shape of the UV-optical non-stellar spectrum.   It is not perfect --- in particular, the increasing flux to shorter wavelengths in the FUV and the dip in the optical spectrum between 4000--5500~\AA\ are not fit --- but illustrates the need to include multiple emission components to describe the broadband spectrum.


\section{Discussion} \label{sec_discussion}

\subsection{The UV Spectrum}

In this manuscript, we present the first FUV spectrum of \vmon.  Due to its quiescent state and moderate reddening along the line of sight to the target, the FUV spectrum is extremely faint, with an observed continuum flux level at $\sim 1\times10^{-17}$~\flam. While the NUV continuum observed with STIS has a red spectral shape, the continuum is flat in the FUV.  Once the spectra are dereddened, the NUV spectral shape is fairly flat but the spectrum begins to turn up to the blue again in the FUV.  The strongest UV emission line is that of \ion{Mg}{2}, but the UV spectra also show broad (FWHM$\sim$2000~km~s$^{-1}$) emission from \ion{Si}{4}, \ion{C}{4}, \ion{He}{2}, and \ion{Fe}{2}. The broad line widths rule out the donor star as the source. The optical Balmer lines are double-peaked, indicating that they originate in the Keplerian accretion disk.  Given the similar line widths of the UV and Balmer lines (which have FWHM of 1900~km~s$^{-1}$ and 2260~km~s$^{-1}$ for H$\alpha$ and H$\beta$, respectively; Marsh, Robinson, \& Wood 1994), it is reasonable to assume that the UV lines also originate in the disk. 

Doublet line fits to \ion{Si}{4} suggest that the line might form in an optically thin gas.  This is in contrast to high accretion rate X-ray binaries in which the lines are optically thick, with accretion disk coronal models predicting optical depths in the lines $\geq10^{4}$  \citep{kallman1991,ko1994}.  The emission lines in the high state have been modeled as originating in an accretion disk corona, a temperature inversion above an optically thick disk induced by X-ray irradiation \citep{ko1994,raymond1993}. However, the  X-ray luminosity is 8 orders of magnitude fainter in quiescence than in outburst for \vmon, and may not be sufficient to generate a disk corona. 


The UV outburst spectrum of the black hole X-ray binary XTE J1118+480 showed an anomalous emission line profile: the \ion{C}{4} and \ion{O}{5} lines were extremely weak while the \ion{N}{5} emission was strong \citet{haswell2002}.  The pattern persists in quiescence, where \ion{N}{5} and \ion{Si}{4} emission lines are detected but \ion{C}{4} is not \citep{mcclintock2003}.  The line ratios are inconsistent with photoionization models wherein \ion{N}{5} emission requires photoionization parameters that also produce \ion{C}{4} and/or \ion{O}{5}. Haswell et al.\ concluded that carbon is underabundant in the system due to CNO processing of material in the donor star. They used the relative nuclear and angular momentum loss time scales in the binary to infer the donor star mass and orbital period at the time the system came into contact. They also predicted that XTE J118+480 is at a later evolutionary stage than \vmon.

The UV spectrum of \vmon\ also shows \ion{C}{4} emission that is weaker than the \ion{Si}{4} emission. \ion{O}{5} 1371~\AA\ is absent in the \vmon\ spectrum, as was also true in XTE J1118+480. Unfortunately, we are not able to look for N enhancements as our chosen COS G140L grating setting did not cover \ion{N}{5} 1240~\AA.  The NIR spectra of \vmon\ also have very weak CO bandhead absorption in the spectrum of the donor star \citep{harrison2007,froning2007}. Comparison of stellar atmosphere models to the NIR spectrum of \vmon\ indicate an underabundance of carbon in the donor star of [C/H]=-1.5. Thus, it is clear from the UV and NIR spectra that \vmon\ is underabundant in C and likely has the same CNO processing mechanism at work that produced the XTE J1118+480 abundance pattern. Haswell et al.'s prediction that XTE J1118+480 is at a later evolutionary stage than \vmon\ is also bourne out by comparisons of UV line ratios:  whereas the \ion{C}{4} line is about 12 times fainter than \ion{Si}{4} or \ion{He}{2} in XTE J1118+480, it is only 1.5 times fainter than \ion{Si}{4} in \vmon\ and actually brighter than \ion{He}{2} (by a factor of 2). Haswell et al.'s  prediction can be tested more quantitatively by obtaining NIR spectra of XTE J1118+480 and modeling the CO bandhead absorption to compare the C abundance to that found for \vmon. 



\subsection{The Broadband SED and Non-Stellar Spectrum Over Time}

\begin{figure}
\includegraphics[angle=90,scale=0.7]{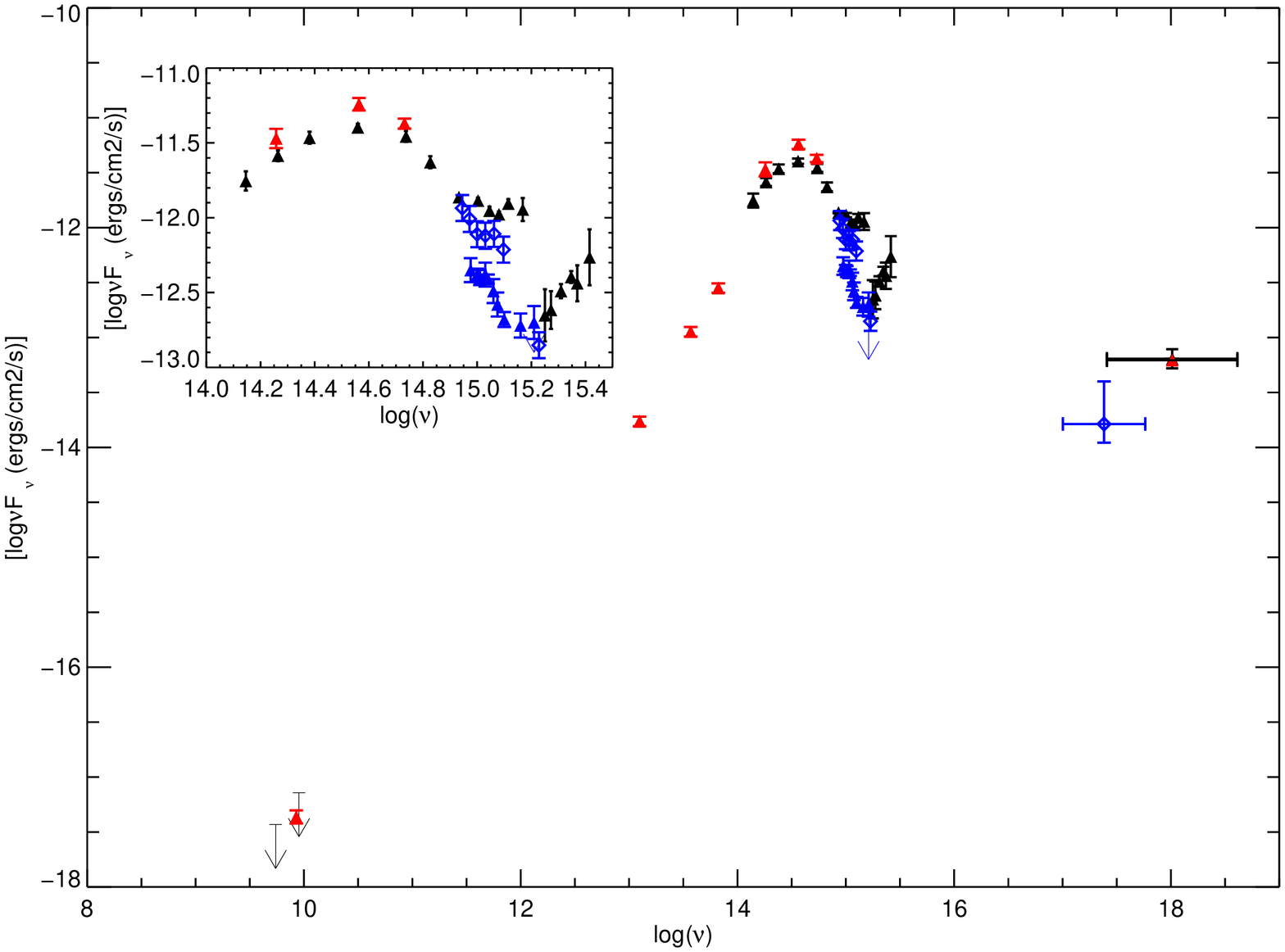}
\figcaption[fig5.ps]{The broadband spectral energy distribution for \vmon. The full SED from radio to X-rays is shown in the main window while the inset gives an expanded view of the NIR/optical/UV range. The solid black triangles are from this work.  The red points show the data from \citet{gallo2007,gallo2006} while the blue points are taken from \citet{narayan1996} (FOS data; open triangles) and \citet{mcclintock2000} (STIS data, closed triangles). Only the data $>$3500~\AA\ is shown from the latter two sources because their points at longer wavelengths have had the donor star contribution removed. All the data have been dereddened using the extinction relation of \citet{cardelli1989} assuming E(B--V) = 0.35.  (Gallo et al.\ originally used a reddening of 0.39 but we have shifted their points to the common value.) \label{fig_sed}}
\end{figure}

\begin{figure}
\includegraphics{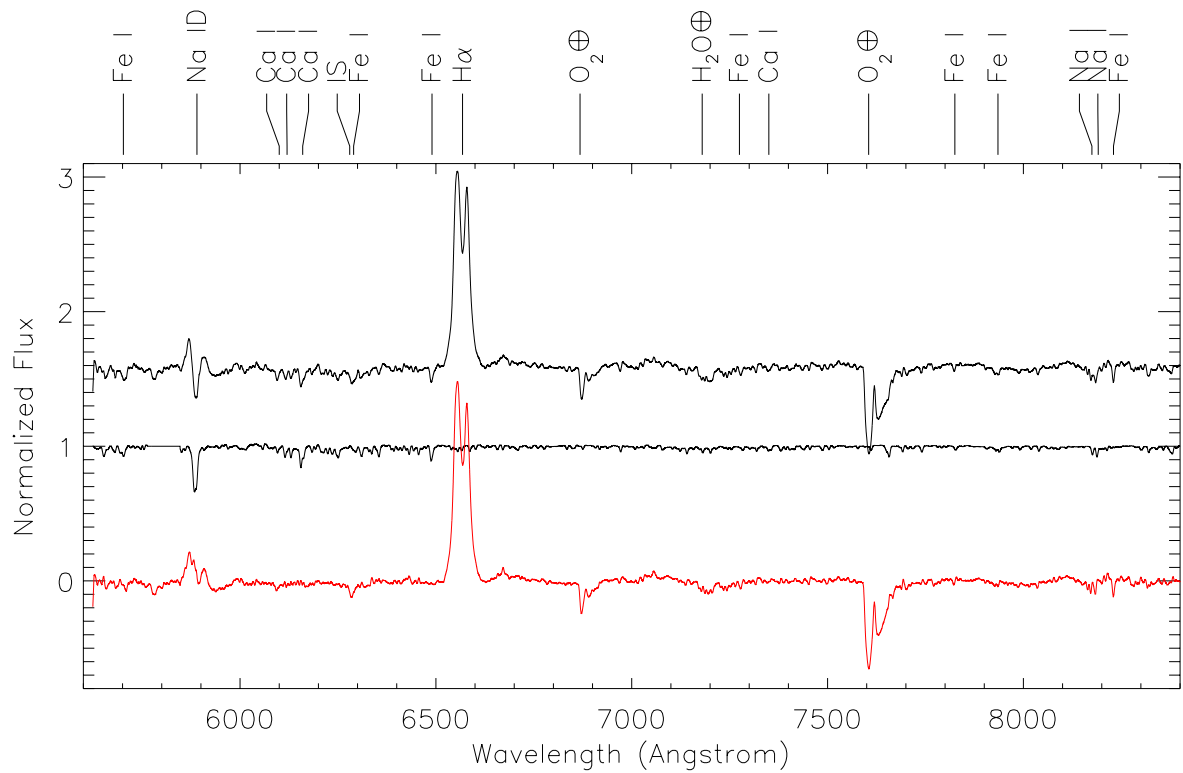}
\figcaption[fig6.eps]{The spectrum of A0620-00 is shown at the top while the Kurucz synthetic spectrum (T=4500~K, $\log$g = 4.5, solar abundances) is shown in the middle. The bottom spectrum in red represents the difference between the spectrum of A0620-00 and the synthetic spectrum when the latter is scaled to 58\% of the observed spectrum.  The feature at 6280~\AA\ marked``IS" is interstellar.  The features marked with circled plus signs are telluric. \label{fig_dilute}}
\end{figure}

\begin{figure}
\includegraphics[scale=0.9]{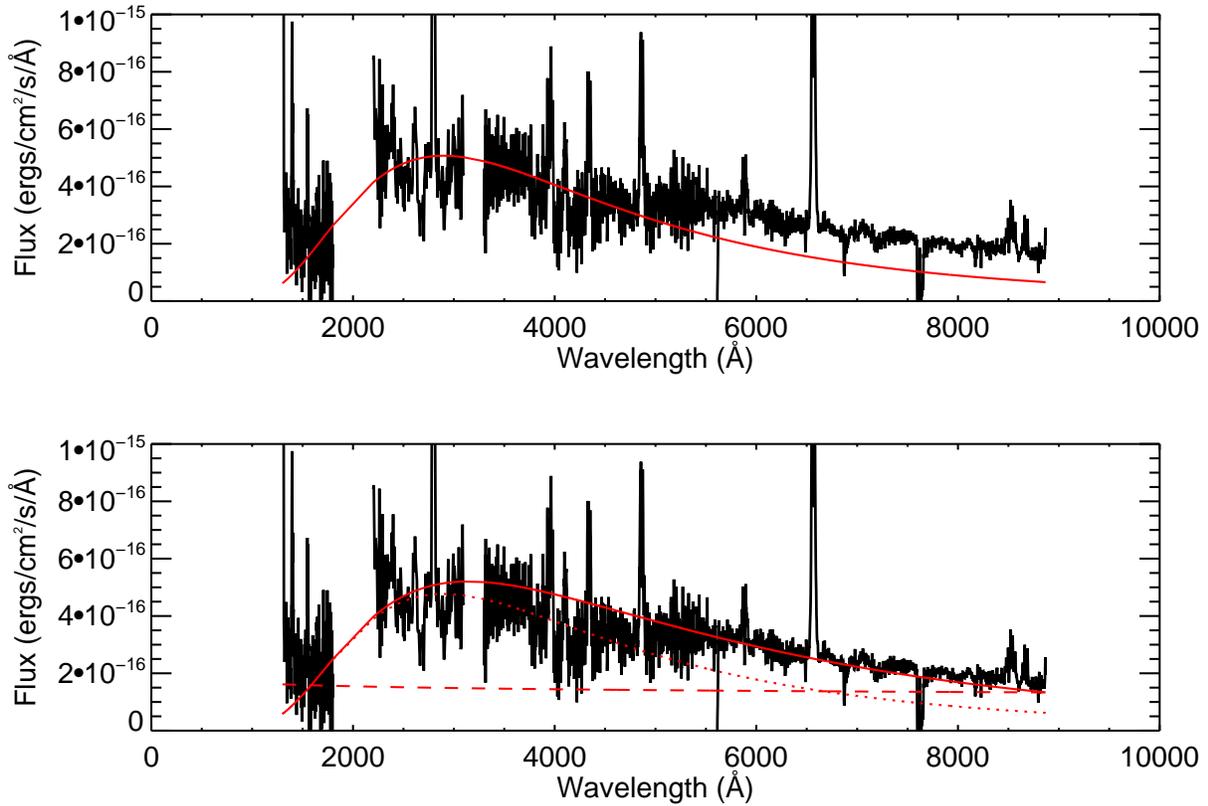}
\figcaption[fig7.ps]{Shown in black in both panels are the dereddened optical and UV spectra of \vmon\ after the donor star contribution has been subtracted. The solid red line in the upper panel shows a 10,000~K blackbody spectrum overplotted.  The lower panel shows the 10,000~K blackbody (dotted red line) and a power law spectrum with $\alpha$ = 1.9 (dashed red line) compared to the observed non-stellar spectrum.  The solid red line is the summation of the blackbody and the power law.\label{fig_disk}}
\end{figure}

Figure~\ref{fig_sed} compares the 2010 SED to two previous SEDs from \citet{mcclintock2000} and \citet{gallo2007}.  The SED of \vmon\ varies over time at all wavelengths for which we have more than one epoch (we have no information for radio or mid-IR variability).  Our optical-NIR data are 20--44\% fainter than those seen by Gallo et al.\ in 2005, while our UV observations are twice as bright as the 1992 FOS data and up to 8 times brighter than the 1998 STIS data.  While the previous UV observations showed a declining intensity at higher energies, the new COS observations reveal a recovery and subsequent upturn in the SED in the FUV.  The relative normalizations of the STIS and COS spectra may be offset due to variability in the target in the day between the observations, but each observation independently supports a UV upturn:  the dereddened NUV spectrum flattens at short wavelengths and the FUV spectrum is blue. The 1998 observations (the solid blue points in Figure~\ref{fig_sed}) also hint at a flattening in the last three NUV points (although the bluest point is an upper limit only).

MR compared the SEDs of \vmon\ and the neutron star X-ray binary, Cen X-4. They particularly emphasized the factor of $\sim$3 drop in the UV intensity in the former while the latter continually rose to the blue. They also noted that the X-ray flux in Cen X-4 was only a factor of $\sim$2 smaller than the UV flux, whereas in \vmon\ the difference is much larger, about an order of magnitude. They interpreted the differing SEDs in light of the ADAF model as evidence of an event horizon in \vmon\ compared to a neutron star surface in Cen X-4.  While the NUV to X-ray ratio for \vmon\ remains large ($\simeq$20) in the latest SED, the FUV flux is only a factor of $\simeq$5 larger than the X-ray value.  Moreover, the blue FUV spectrum suggests a recovery in the flux to shorter wavelengths rather than a continuous drop to the blue.  Thus, the difference in the UV SEDs between the black hole and neutron star systems is less dramatic at present than at the time of the MR comparison. In their study of three black hole and one neutron star X-ray binaries, \citet{hynes2011} also note that the NUV SEDs do not differ noticeably between the two types of systems, although the X-ray to NUV luminosity ratio is always higher in the neutron star binaries.

To constrain the source(s) of variability in \vmon, we constructed a broadband UV-optical spectrum of the system after the contribution of the donor star is subtracted (Figure~\ref{fig_disk}). In that spectrum, the peak emission in the non-stellar component occurs near 3000~\AA.  A blackbody fit to the peak gives a temperature of T$\simeq$9000--11000~K.  This is very similar to the 9000~K blackbody that \citet{mcclintock1995} fit to the FOS observations of \vmon, though our source must be larger and/or hotter to match the higher observed flux in 2010. A temperature increase is the likely cause, given that our spectrum is also more blue.  Indeed, the 10,000~K blackbody in the upper panel of Figure~\ref{fig_disk} corresponds to an emitting area of $\pi (0.09 R_{\odot})^2$, equivalent to the emitting area seen in 1992.  Thus, assuming that a single thermal component is present in the quiescent \vmon, it appears to be fairly stable, with small-scale temperature and/or emitting area changes over the past two decades. As noted by \citet{mcclintock1995}, the emitting area is too small to correspond to the full accretion disk.  The UV/optical SEDS of several other X-ray binaries show evidence of blackbody components at comparable size and temperature as our results. For example, \citet{park2011b} found a blackbody component with $T\sim13000$~K in Cen X-4. The radius, $\sim2\times10^{9}$~cm is comparable to ours, $\sim4.5\times10^{9}$~cm.  Similarly, \citet{hynes2011} has fit the SEDS of four LMXBs and found blackbody emission at T = 5000 -- 13,000~K in three of them.  All have emitting areas much smaller than the area of the accretion disk. 

There are several potential sources for this component.  Perhaps the most likely is the bright spot (the accretion stream-disk impact point), which is seen in quiescent Doppler tomographic maps of the accretion disk in \vmon\ \citep{marsh1994, shahbaz2004,neilsen2008}. Other possibilities include: a) the bright spot is located at a smaller radius in the disk than the circularization radius and the corresponding \.{m} is lower (although this is inconsistent with the bright spot position in the outer disk in optical Doppler maps); b) the thermal source is closer to the center of the system than the bright spot, such as the transition radius in the inner disk \citep{hynes2011}; c) the emission source is optically thin emission from the disk (though Hynes \& Robinson note that a significant Balmer jump should be present in that case, which is not seen in the optical spectrum of \vmon); or d) the mass transfer rate we measure is currently higher than the average rate indicated by the interoutburst interval (though the mass transfer rate seen in the 1992 FOS observations was also a factor of several above that value). 

For \vmon, we favor the bright spot as the source, given that it is directly observed in the optical Doppler maps of the system. If we assume that the bright spot is the blackbody source, we can 
estimate the mass accretion rate following the method of \citet{park2011b} and equating the luminosity of the spot with the blackbody luminosity: 
\begin{equation}
L_{BS} = G M_{BH} \mbox{\.{m}} \left (\frac{2}{R_{circ}} - \frac{1}{R_{L_{1}}} \right )  = 4 \pi R^{2}_{BB}\sigma T^{4}_{BB}
\end{equation}
where $L_{BS}$ is the maximum luminosity of the bright spot \citep{menou2001} and $R_{BB}$ and $T_{BB}$ are the blackbody radius and temperature we derive above.  System parameters (masses, mass ratios, orbital period) are taken from \citet{cantrell2010}. From this, we obtain a mass accretion rate at the bright spot of \.{m} = 3.4$\times$10$^{-10}$~\msun~yr$^{-1}$.  This rate is substantially larger than the  \.{m} $< 5\times10^{-15}$~\msun~yr$^{-1}$ rate at the black hole inferred by the X-ray luminosity \citep{mcclintock1995}. McClintock et al.\ attribute the discrepancy to inefficient mass transfer through the quiescent disk within the paradigm of the disk instability model. However, the disk mass transfer rate inferred from the interval between the two observed outbursts of \vmon\ gives  \.{m} $\sim 3 \times10^{-11}$~\msun~yr$^{-1}$ \citep{mcclintock1983}, which is an order of magnitude below the rate we infer for the bright spot. To reconcile these last two numbers requires that $\sim$90\% of the mass flowing through the bright spot must be lost from the disk if the disk instability model and the 58 yr outburst recurrence interval for \vmon\ are correct.  The relative mass transfer rates inferred from the X-ray luminosity and the bright spot luminosity require that virtually all of the accreted mass fails to reach the black hole and/or falls into the black hole without radiating efficiently, as through an ADAF. Winds and/or a jet may carry off much of the material.   The original ADAF models noted the likelihood of outflows linked with ADAFs \citep{narayan1995}, and \citet{blandford1999}, in their ``ADIOS'' model, propose that only a small fraction of the accreted material falls into the black hole.

Finally, we note that the broad spectral coverage in our observations shows that a single blackbody does not describe the full UV-optical spectrum. The addition of a powerlaw component does a reasonable job of improving the fit at long wavelengths, though the overall fit still deviates from the observed spectrum in the FUV and in parts of the optical spectrum (most notably between 4000--5000~\AA).  The powerlaw is not a unique description but it does illustrate the need for a second source in addition to the thermal component to match the red optical spectrum. The powerlaw is fairly flat in the optical but is not steep enough to match the FUV upturn, which suggests the need for a more complex, physically-based model for the broadband spectrum. \citet{park2011} note that a multicomponent model may also be needed in Cen X-4 to explain changes in the NUV to X-ray flux ratio in the SED over time. In the following section, we compare our observations to two published models of \vmon in quiescence, the ADAF and maximally-jet dominated models.


\subsection{SED Comparison to Quiescent Models}

\citet{narayan1997} fit an updated advection-dominated accretion flow (ADAF) model to the SED of \vmon\ using data first presented in \citet{narayan1996}.  The SED consisted of the UV/optical FOS data discussed above as well as a ROSAT X-ray point and optical data from the literature.  Our NUV data are slightly brighter than the ADAF model (which was fit to the fainter FOS data) but the shape of the model and our NUV data are consistent.  However, the ADAF model strongly underpredicts the FUV flux (by a factor $\sim$6) and the blue spectral shape in that region.  Comparing our data to the variations in model values presented in \citet{narayan1996} (Figures 6 \& 7) suggests that the two would be in closer agreement for those models in which the transition radius from thin disk to ADAF, r$_{tr}$, is decreased from $\log r_{tr} = 3.8$ to $\log r_{tr} = 3.0$. Decreasing the inner radius of the disk unsurprisingly has the effect of shifting the peak in the disk model SED to the blue and increasing the FUV emission. The smaller transition radius worsens the fit to the fainter 1992 FOS data, however, which is entirely inconsistent with the $\log r_{tr} = 3.0$ model.  Thus, viewed within the ADAF picture, the time-variable SED suggests changes in the structure of the disk and the ADAF during quiescence, including possible changes in the size of the ADAF radius by a factor of 6. 

Changing the disk transition radius alone will not reconcile the ADAF model to our observations, however.  The decrease in the transition radius does not alter the predicted X-ray fluxes in the model, whereas the X-rays also brightened in the 2010 data compared to the earlier measurements, indicating the need to change other model parameters to match the data. Ideally, a new ADAF model fit that takes into account the full SED would be performed. The published ADAF model of \vmon\ is over a decade old and so does not reflect many of the recent changes in the model.  Given the discrepancies between the mass accretion rate inferred at the bright spot and lower rates inferred from the interoutburst interval and the X-ray luminosity in \vmon, a model that incorporates significant mass loss may be indicated. More recent ADAF models of other targets add an outflow component; e.g., \citet{yuan2005}.  The outflow in that model was motivated by the radio detections of low-state XRBs, but the outflow can also contribute synchrotron emission at shorter wavelengths (e.g., Figure 2 of Yuan et al.).  Thus, the addition of outflow components can potentially reconcile the ADAF model with the excess FUV emission seen in \vmon.

\citet{gallo2007} fit a maximally-jet dominated model to the broadband spectrum of \vmon.  We compared our SED to their model (Figure 4 in their paper) and found good agreement between their model and our UV measurements.  In particular, the observed upturn in flux from the NUV to the FUV is consistent with the model shape in this region, as the donor star and disk and the pre-acceleration inner jet synchrotron components drop rapidly and the post-acceleration outer jet synchrotron component becomes the dominant emitter.  We examined in more detail how the maximally-jet dominated model would change when fit to the 2010 data by running new fits including our observations.  Because we had limited radio and X-ray data, we fit a hybrid data set, taking the radio, mid-IR, and X-ray data from the \citet{gallo2007} epoch and the NIR, optical, and UV from our observations. The primary result when comparing the jet-dominated model to the 2010 data is that a thermal component does not account for the excess emission in the FUV.  In the model the thermal emission comes from a cool donor star plus a multi-temperature blackbody disk and peaks in the I band, after which it drops rapidly, contributing negligible emission in the FUV.  Instead, the dominant source for the FUV fluxes is non-thermal synchrotron emission from the jet. The pre-acceleration inner jet component of the model dominates the FUV (and NUV shortward of $\simeq$3100~\AA), although there can also be a contribution (about 30\% in our fit) from the post-acceleration outer jet component.  Note that although we did not detect radio emission from \vmon, our upper limits from the ATCA were larger than the previous detected VLA fluxes, so the jet may have been present at levels below our detection threshold.

For both models, the question of the treatment of the thin disk component must be reexamined. The thin disk in both cases is modeled using a multi-temperature blackbody.  The models do not include irradiation of the outer disk and donor star or a bright spot at the accretion stream impact point. Irradiation is not significant for \vmon\ in quiescence given its low X-ray luminosity \citep{vanparadijs1984}, but the treatment of the accretion disk and bright spot affects interpretation of the SED in the optical-UV in particular. It is not clear that a thermal, steady-state accretion disk is the right model for a quiescent soft X-ray transient disk. In cataclysmic variables, eclipse mapping has long indicated that the accretion disks of quiescent dwaf novae have flat brightness temperature profiles that do not match the $T(R) \propto R^{-3/4}$ profile expected for a steady-state disk \citep{horne1993}. The $\simeq$10,000~K thermal component we fit to our observations is too small to originate in the bulk of the disk; the hot spot is more likely if a thermal source is present. 


\section{Conclusions} 

We have presented broadband observations of the black hole X-ray binary, \vmon, centered around the first FUV spectroscopy of the system.  Our primary results are as follows:

\begin{enumerate}

\item The observed spectrum of \vmon\ is red in the NUV and flat at FUV wavelengths.  The dereddened spectra show a flat spectrum in the NUV and a steady increase to the blue in the FUV. The spectra show prominent, broad (FWHM$\sim$2000~km~s$^{-1}$) emission lines of \ion{Si}{4}, \ion{C}{4}, \ion{He}{2}, \ion{Fe}{2}, and \ion{Mg}{2}.  The \ion{C}{4} line is anomalously weak, which is consistent with the weak C abundance seen in NIR spectra of the donor star.  The relative strength of the \ion{C}{4} line is not as low as that in XTE J1118+480, consistent with the predictions of \citet{haswell2002} that the latter is at a later stage of binary evolution.

\item Comparisons with previous NUV spectroscopy of \vmon\ show that it is highly variable at these wavelengths, with the most recent data being up to 8 times brighter than previous observations. Comparison of our data from night to night suggests that \vmon\ can also vary by $\sim$50\% on short time scales.

\item We constructed a broadband (radio through X-ray), dereddened spectral energy distribution of \vmon, based on semi-contemporaneous data acquired over a four-day period. Comparison of the SED with two previous ones shows that the system varies at all wavelengths for which we have multiple samples.  The new data reveal that the continuous decrease in flux to shorter wavelengths seen in previous optical-UV data does not continue into the FUV, which instead shows a blue upturn. Comparison between 2005 and 2010 observations show variations of up to 44\% in the optical/NIR with no change in X-ray flux.

\item The UV-optical spectrum with the donor star contribution removed shows a peak near 3000~\AA. A single blackbody with $T\simeq10000$~K fits the peak and has the same emitting area as a 9000~K blackbody fit to an earlier observation by \citet{mcclintock1995}. The emitting area is too small to be consistent with the bulk of the disk emitting as a thermal blackbody.  A possible thermal source is the mass accretion stream-disk impact point (the ``bright spot''). 

\item The single blackbody component does not match the full UV-optical spectrum, in particular the FUV upturn and the optical flux longward of 5000~\AA. The addition of a powerlaw with $\alpha$=1.9 provides a qualitatively improved fit to the spectral shape but continued deviations suggest the need for more sophisticated, physically-based model fits to the data.

\item By assuming that the blackbody component is emitted by the hot spot at the disk edge, we calculated a mass accretion rate from the hot spot luminosity.  This accretion rate is an order of magnitude larger than the rate inferred from the interoutburst interval and the disk instability model.  To reconcile these requires that $\sim$90\% of the accreted material be lost from the system. The accretion rate at the hot spot is $10^{5}$ the accretion rate at the black hole as inferred from the X-ray luminosity. This indicates that virtually all of the accreted material must escape the system, remain in the thin disk, and/or be radiatively inefficient in the inner region of the system.

\item Comparisons to a previous ADAF model of \vmon\ shows that the transition radius between the thin disk and the ADAF may need to be decreased to match the brighter UV observations in the recent observations and may indicate changes in the relative disk/ADAF sizes over time.  Alternately, a revised model that includes mass loss in the form of winds or a jet may reconcile the ADAF model with the observed SED. Comparisons to maximally jet-dominated models indicate that the UV emission shortward of 3100~\AA\ is dominated by non-thermal synchrotron emission from the jet. 

\end{enumerate}

\acknowledgments
Thanks to Amanda Bayless for setup work on the ground-based observations and to Derck Massa for useful discussions on interstellar reddening. Based on observations made with the NASA/ESA Hubble Space Telescope, which is operated by the Association of Universities for Research in Astronomy, Inc., under NASA contract NAS 5-26555. These observations are associated with program 11550. This work was supported by NASA grant NNX08AC146 to the University of Colorado at Boulder. Some of the data presented in this paper were obtained from the Multimission Archive at the Space Telescope Science Institute (MAST). Support for MAST for non-HST data is provided by the NASA Office of Space Science via grant NNX09AF08G and by other grants and contracts. The Australia Telescope is funded by the Commonwealth of Australia for operation as a National Facility managed by CSIRO. 


{\it Facility:} \facility{HST (COS, STIS), SWIFT, CTIO:1.3m, ATCA, KECK}

\clearpage
\begin{deluxetable}{lccccc}
\tablecaption{Observation Summary \label{tab_obs}}
\tablewidth{0pt}
\tablecolumns{6}
\tablehead{
\colhead{Telescope} & \colhead{Instrument} & \colhead{Grating/Filter} & \colhead{Date (UT)} & \colhead{Time (UT)} & \colhead{T$_{exp}$ (sec)}  }
\startdata
HST & COS & G140L &  2010 Mar 23 & 20:36 & 16676  \\
HST & STIS &  G230L & 2010 Mar 24 &  20:45 & 10472  \\
SWIFT & XRT & & 2010 Mar 22 & 04:50\tablenotemark{a} & 3772 \\
SWIFT & UVOT & UVW2 & 2010 Mar 22 & 04:50 &  1228 \\
SWIFT & XRT &  & 2010 Mar 23 & 04:58 & 3894 \\
SWIFT & UVOT & UVM2 & 2010 Mar 23 & 04:58 &  1335 \\
SWIFT & XRT &  & 2010 Mar 24 & 00:12 & 5295 \\
SWIFT & UVOT & UVW1 & 2010 Mar 24 & 00:12 &  803 \\
SWIFT & XRT &  & 2010 Mar 25 & 03:29 & 5378 \\
SWIFT & UVOT & U & 2010 Mar 25 & 03:29 &  851 \\
CTIO 1.3-m & ANDICAM & BVIJHK & 2010 Mar 18 -- 31\tablenotemark{b} & \nodata & \nodata \\
Keck I & LRIS &  600/7500, 600/4000 & 2010 Mar 25 & 05:37 & 600 \\
 ATCA & CABB & 5.5 and 9 GHz & 2010 Mar 23 & 06:15 & 12816 \\
 ATCA & CABB & 5.5 and 9 GHz & 2010 Mar 24 & 06:30 & 12384 \\
\enddata
\tablenotetext{a}{We list the start time of each Swift pointing in a band, but due to Swift observation algorithms, the actual data acquisition is typically spread out over several hours.}
\tablenotetext{b}{Weather permitting, the SMARTS observations were acquired nightly between 0:00 and 3:00 UT. }
\end{deluxetable}

\clearpage
\begin{deluxetable}{lcc}
\tablecaption{UV Line Fluxes \label{tab_lines}}
\tablewidth{0pt}
\tablecolumns{3}
\tablehead{
\colhead{Line} & \colhead{Line Flux} & \colhead{FWHM} \\
 (\AA) & (10$^{-16}$ ergs~cm$^{-2}$~s$^{-1}$) & (km~s$^{-1}$) }
\startdata
Si~\sc{IV} 1393.8 & 3.7$\pm$0.5 & 2168$\pm$526 \\
Si~\sc{IV} 1402.8 & 1.8 & 2168$\pm$526 \\
C~\sc{IV}\tablenotemark{a} 1548.2 & 1.8$\pm$0.3 & 2034$\pm$326 \\
C~\sc{IV} 1550.8 & 1.8 & 2034$\pm$326 \\
He~\sc{II} 1640.4 & 1.8$\pm$0.1 & 1477$\pm$258 \\
Fe~\sc{II} 2609 & 8.8$\pm$0.2 & 3095$\pm$566 \\
Mg~\sc{II} 2796.4/2803.5\tablenotemark{b} & 91$\pm$2 & 2438$\pm$61 \\
\enddata
\tablenotetext{a}{The \ion{C}{4} doublets FWHM were fixed to a 1:1 ratio.}
\tablenotetext{b}{The \ion{Mg}{2} doublet was fit with a single component.}
\tablecomments{Fluxes are observed values, uncorrected for reddening.}
\end{deluxetable}

\clearpage
\begin{deluxetable}{lccccc}
\tablecaption{UV Continuum Fluxes \label{tab_cont}}
\tablewidth{0pt}
\tablecolumns{6}
\tablehead{
\colhead{Date} & \colhead{Instrument} & \colhead{Filter} & \colhead{$\lambda_{c}$} & \colhead{Width\tablenotemark{a}} & \colhead{F$_{\lambda}$} \\
& & & (\AA) & (\AA) & (10$^{-17}$ ergs~cm$^{-2}$~s$^{-1}$~\AA$^{-1}$)  }
\startdata
23 Mar & COS & \nodata & 1112.6 & 23.6 & 2.7$\pm$1.6 \\
23 Mar & COS & \nodata & 1152.5 & 24.7 & 1.00$\pm$0.43 \\
23 Mar & COS & \nodata & 1280.1 & 19.1 & 1.16$\pm$0.32 \\
23 Mar & COS & \nodata & 1347.5 & 64.1 & 1.53$\pm$0.14 \\
23 Mar & COS & \nodata & 1469.6 & 99.0 & 1.47$\pm$0.16 \\
23 Mar & COS & \nodata & 1600.0 & 58.5 & 1.17$\pm$0.34 \\
23 Mar & COS & \nodata & 1684.9 & 68.6 & 1.07$\pm$0.43 \\
24 Mar & STIS & \nodata & 2034.0 & 167.0 & 2.94$\pm$0.52 \\
24 Mar & STIS & \nodata & 2300.1 & 198.1 & 3.10$\pm$0.21 \\
24 Mar & STIS & \nodata & 2490.5 & 176.6 & 4.11$\pm$0.18 \\
24 Mar & STIS & \nodata & 2699.7 & 99.2 & 5.39$\pm$0.33 \\
24 Mar & STIS & \nodata & 2975.8 &  248.3 & 7.01$\pm$0.32 \\
22 Mar & UVOT & UVW2 & 2030 & 760 & 11.8$\pm$1.9\tablenotemark{b} \\
23 Mar & UVOT & UVM2 & 2231 & 510 & 5.7$\pm$1.7 \\
24 Mar & UVOT & UVW1 & 2634 & 700 & 10.1$\pm$1.8 \\
25 Mar & UVOT & U & 3501 & 875 & 7.9$\pm$1.6 \\
\enddata
\tablenotetext{a}{For the COS and STIS data, the width is the size of the wavelength range (centered on $\lambda_{c}$) over which the continuum flux average was calculated. For UVOT, width refers to the FWHM of the imaging filter used.}
\tablenotetext{b}{Exposure marred by a detector feature running through the object location.}
\tablecomments{Fluxes are observed values, uncorrected for reddening.}
\end{deluxetable}

\clearpage
\begin{deluxetable}{cccc}
\tablecaption{Spectral Energy Distribution \label{tab_sed}}
\tablewidth{0pt}
\tablecolumns{4}
\tablehead{
\colhead{Band} & \colhead{Instrument} & \colhead{$\log(\nu)$} & \colhead{$\log(\nu F_{\nu}$)}  \\
& & (Hz) & (ergs~cm$^{-2}$~s$^{-1}$)}
\startdata
X-ray & Swift & 18.0118 & -13.1938 \\
FUV & COS & 15.4152 & -13.9384 \\
FUV & COS & 15.3696 & -13.8283 \\
FUV & COS & 15.3473 & -13.6858 \\
FUV & COS & 15.3096 & -13.6655 \\
FUV & COS & 15.2727 & -13.7277 \\
FUV & COS & 15.2503 & -13.7440 \\
NUV & STIS & 15.1685 & -13.2233 \\
NUV & STIS & 15.1151 & -13.1469 \\
NUV & STIS & 15.0805 & -12.9899 \\
NUV & STIS & 15.0455 & -12.8371 \\
NUV & STIS & 15.0032 & -12.6807 \\
U & UVOT & 14.9326 & -12.5593 \\
B & ANDICAM & 14.8270 & -12.1950 \\
V & ANDICAM & 14.7372 & -11.8910 \\
I & ANDICAM & 14.5593 & -11.6420 \\
J & ANDICAM & 14.3811 & -11.5882 \\
H & ANDICAM & 14.2643 & -11.6653 \\
K & ANDICAM & 14.1459 & -11.8057 \\
Radio &  ATCA & 9.7388 & -17.4319\tablenotemark{b} \\
Radio &  ATCA & 9.9542 & -17.1417\tablenotemark{b} \\
\enddata
\tablenotetext{a}{Assumes N$_{H}$ = 1.6$\times10^{21}$~cm$^{-2}$. }
\tablenotetext{b}{Upper limits.}
\tablecomments{With the exception of the X-ray point, fluxes are observed values, uncorrected for interstellar extinction.}
\end{deluxetable}

\clearpage
\pagestyle{empty}

\clearpage
\pagestyle{empty}


\clearpage
\pagestyle{empty}

\clearpage
\pagestyle{empty}

\clearpage
\pagestyle{empty}

\clearpage
\pagestyle{empty}

\clearpage
\pagestyle{empty}

\end{document}